\documentclass[conference]{IEEEtran}

\usepackage{cite}
\usepackage{amsmath,amssymb,amsfonts}
\usepackage{algorithmic}
\usepackage{graphicx}
\usepackage{textcomp}
\usepackage{xcolor}
\usepackage{bm}
\usepackage{upgreek}

\usepackage{graphicx}
\usepackage{subcaption}
\usepackage[ruled,linesnumbered]{algorithm2e}
\usepackage{amsmath}

\def\BibTeX{{\rm B\kern-.05em{\sc i\kern-.025em b}\kern-.08em
    T\kern-.1667em\lower.7ex\hbox{E}\kern-.125emX}}

\begin{document}

\title{End-to-End Delay Minimization based on Joint Optimization of DNN Partitioning and Resource Allocation for Cooperative Edge Inference\\}

\author{
    \IEEEauthorblockN{Xinrui Ye$^{\dagger}$, Yanzan Sun$^{\dagger}$, Dingzhu Wen$^{\ddag}$, Guanjin Pan$^{\dagger}$, Shunqing Zhang$^{\dagger}$}
    \IEEEauthorblockA{ $^{\dagger}$ Key laboratory of Specialty Fiber Optics and Optical Access Networks, Shanghai University, Shanghai, 200444, China\\
    $^{\ddag}$ School of Information Science and Technology, ShanghaiTech University, Shanghai, 200031, China}
    }

\maketitle

\begin{abstract}
Cooperative inference in Mobile Edge Computing (MEC), achieved by deploying partitioned Deep Neural Network (DNN) models between resource-constrained user equipments (UEs) and edge servers (ESs), has emerged as a promising paradigm. Firstly, we consider scenarios of continuous Artificial Intelligence (AI) task arrivals, like the object detection for video streams, and utilize a serial queuing model for the accurate evaluation of End-to-End (E2E) delay in cooperative edge inference. Secondly, to enhance the long-term performance of inference systems, we formulate a multi-slot stochastic E2E delay optimization problem that jointly considers model partitioning and multi-dimensional resource allocation. Finally, to solve this problem, we introduce a Lyapunov-guided Multi-Dimensional Optimization algorithm (LyMDO) that decouples the original problem into per-slot deterministic problems, where Deep Reinforcement Learning (DRL) and convex optimization are used for joint optimization of partitioning decisions and complementary resource allocation. Simulation results show that our approach effectively improves E2E delay while balancing long-term resource constraints.
\end{abstract}


\section{Introduction}

The field of Artificial Intelligence (AI) has experienced substantial advancements in recent years, especially in areas such as Automatic Speech Recognition (ASR), Natural Language Processing (NLP), and Computer Vision (CV). The incorporation of AI across various sectors like Augmented Reality (AR), autonomous vehicles, and drones, has heightened significantly, improving terminal devices' perception of the environment and consequently enriching the user's interactive experience \cite{6G}.


AI tasks, including but not limited to speech recognition, pedestrian detection, and lane recognition, typically necessitate low delay. At the same time, the requirement for extended standby durations in terminal devices such as AR glasses and mobile phones compels efforts toward reducing terminal energy consumption \cite{6G}. Nevertheless, the computational complexity of AI algorithms, particularly Deep Neural Network (DNN), remains high \cite{complex}. Therefore, conducting DNN inference on resource-limited mobile devices in a prompt and reliable fashion poses a significant challenge.


To mitigate this issue, the concept of edge AI \cite{letaief-6G} has been introduced. Edge AI leverages Mobile Edge Computing (MEC) to aid with AI training and inference. With the powerful computing capabilities of MEC, edge AI significantly reduces system delay and energy consumption. 

Some approaches either complete the entire DNN inference task locally or offload it to the edge \cite{guangjin}, \cite{xuemin-accuracy}. However, this methodology cannot exploit the full computational resources of both local devices and edge servers \cite{xuemin-accuracy}. An alternative proposal involves model partitioning, which disaggregates the model by layers. Each layer can be inferred locally or at the edge, thereby reducing inference delay \cite{neurosurgeon}. Specifically, the Edgent framework is designed to simultaneously optimize DNN partitioning and DNN right-sizing, with the aim of maximizing inference accuracy \cite{edgent}. The DNN surgery strategy optimizes the E2E delay under dynamic workloads by applying DNN partitioning for continuous task arrivals \cite{dnnsurgery}. Nonetheless, these solutions only take into account DNN offloading optimization for a single user.

A multi-user context is considered in \cite{Multi-exit}, which combines DNN early exiting with partitioning to expedite inference while maintaining accuracy. A separate study \cite{energy} explores the trade-off between inference efficiency and energy consumption in heterogeneous computing resources. However, neither of these studies accounts for the queuing delay caused by partitioned offloading under multi-task workloads, nor do they consider delay performance from a long-term perspective.

In this work, we consider a multi-user MEC system, wherein users are engaged in different AI inference tasks. These AI models are also deployed on an ES. The main contributions of this paper are two-fold:

\begin{itemize}
   \item{\textit{E2E Delay Minimization Formulation based on Serial Queue Model.}} We employ a serial queue model based on queuing theory to obtain the E2E delay experienced by task processes after DNN partitioning. Based on this model, we formulate the edge collaborative inference problem as a long-term E2E delay minimization problem for inference tasks, while adhering to energy, communication, memory, and computing resource constraints.
   
  \item{\textit{Lyapunov-guided Multi-Dimensional Optimization (LyMDO) algorithm.}} We develop a Lyapunov-guided approach that combines DRL with convex optimization. Initially, we leverage Lyapunov optimization to transform the multi-slot stochastic problem into per-slot deterministic problems, which can be solved individually in each time slot. Subsequently, DRL is employed to ascertain the layer-level DNN partitioning decision, with further support from convex optimization to handle the multi-resource allocation problem. The proposed integrated approach considerably enhances the convergence speed of DRL.
  
\end{itemize}

Moreover, we compare the proposed LyMDO algorithm with DRL-based baseline algorithms. Experimental results demonstrate that LyMDO outperforms these baseline algorithms by reducing delay by approximately 30\%.

\begin{figure}
\centering 
\includegraphics[height=1.8in, width=2.563in]{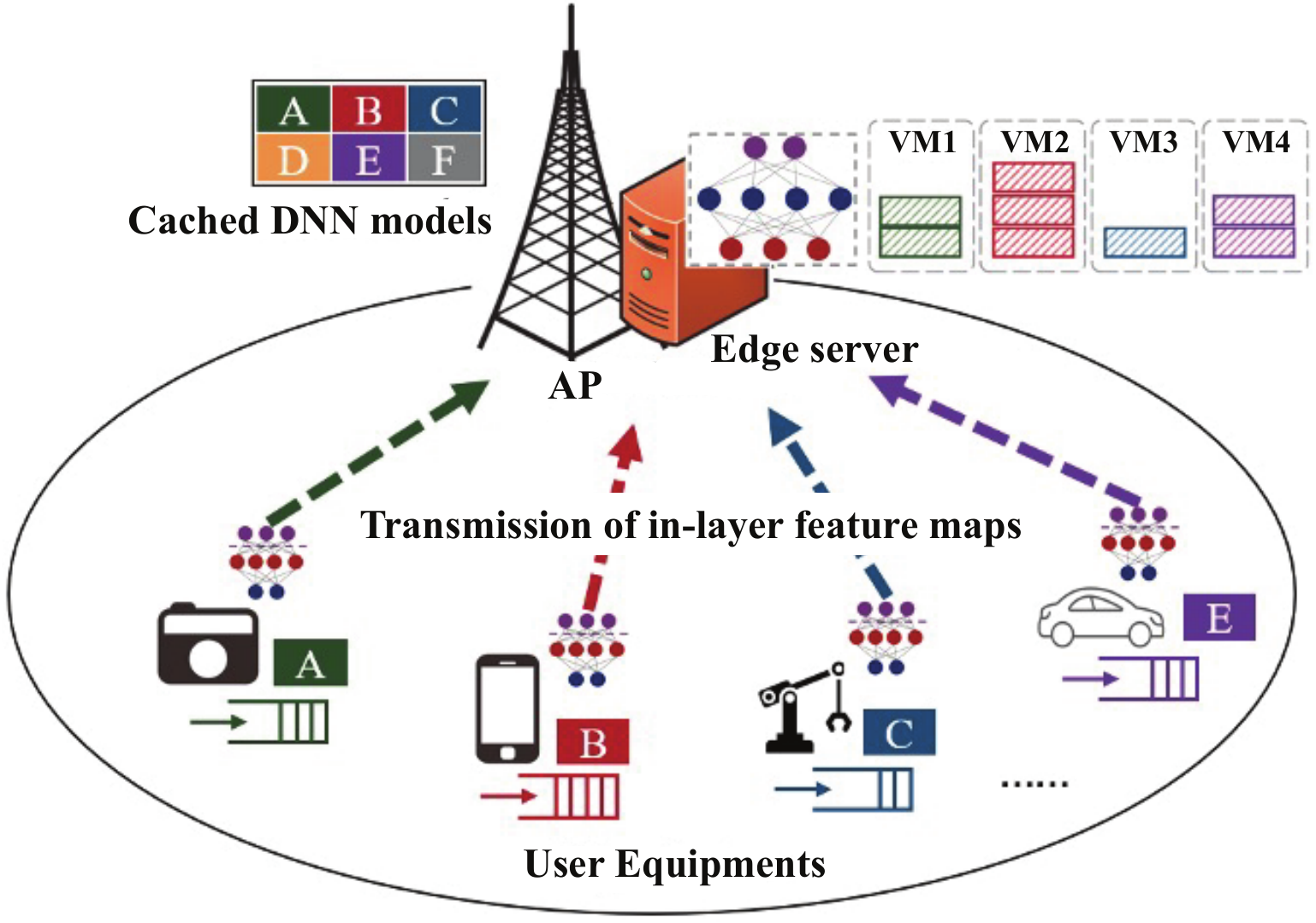} 
\caption{Cooperative inference of DNN partitioning in MEC.} 
\label{system model} 
\end{figure} 

\addtolength{\topmargin}{0.011in}

\section{System Model}
In this section, we introduce a collaborative edge inference system based on DNN partitioning within a single-cell MEC scenario. As shown in Fig. 1, the system involves $N$ UEs, collectively represented by the set $\mathcal{N}=\{1,2, \ldots, N\}$, each performing AI tasks $\phi_n$. To accelerate the execution of the DNN inference task, each UE possesses the capability to make the layer-level offloading decision with the edge server (ES), a mechanism known as DNN partitioning \cite{neurosurgeon}. We divide the time domain into multiple discrete time slots, defined as $\mathcal{T}=\{1,2, \ldots, K\}$, and each slot $t\in \mathcal{T}$ has equal duration $\tau$.



\subsection{DNN Partitioning Model}
To improve the generalizability of our modeling to various DNN architectures, we employ sequential computational graph abstraction. We classify both linear layers (such as the convolutional layer in AlexNet) and parallel layers (like the recurrent execution units in ResNet) as logical layers, similar to previous work \cite{2020joint}.


For DNN $\phi_n$, let $\mathcal{L}_n=\{1,2, \ldots, L_n\}$ denote the sequence of logical layers. For each layer $l\in\mathcal{L}_n$, we use $M_n(l)$ to denote the number of requested multiply-and-accumulate operations (MACs) for executing this layer \cite{guangjin}. The memory required to load layer $l$ is represented by $C_n(l)$, and $\psi_n(l)$ corresponds to the data size of the output feature map. In each slot $t$, an offloading decision (also known as a partitioning decision) of AI tasks $\phi_n$ can be modeled with an integer variable $\ell_n^t\in\mathcal{L}_n$ indicating that the pre-trained DNN model of layers 0 to $\ell_n^t$ are loaded and executed locally while the remaining pre-trained DNN model, layers $\ell_n^t+1$ to $L_n$, are loaded and executed by ES’s corresponding virtual machine (VM) after the transmission of in-layer feature maps $\psi_n\left(\ell_n^t\right)$.

\subsection{E2E delay Model}
The E2E delay of executing a DNN inference task comprises three components: the sojourn delay on the UE and ES’s VM, and the transmission delay of feature maps. Generally, we neglect the return delay due to the small data size of its result. Therefore, the average E2E delay of $\phi_n$ can be written as
\begin{align}
T_n^{t, E 2 E}=T_n^{t, u e}+T_n^{t, t r a n s}+T_n^{t, e s}
\end{align}
where $T_n^{t, u e}$ is the local sojourn delay, $T_n^{t, t r a n s}$ is the transmission delay, and $T_n^{t, e s}$ is the edge sojourn delay.

The arrivals of task $\phi_n$ on UE $n$ follows a Poisson distribution with an arrival rate of $\lambda_n^t$ in slot $t$. We present three components model of delay below.

1) Local sojourn delay: The task queue of UE $n$ accepts task arrivals and uses its computational resource $f_n^{t, u e}$ (in CPU cycle/s) to perform the local inference of the task's designated portion. Based on M/D/1 queueing system. the average local sojourn delay of a task $\phi_n$ in UE $n$ can be calculated as \cite{queueing},
\begin{align}
T_n^{t, u e}=\frac{1}{\mu_n^{t, u e}}+\frac{\lambda_n^t}{2\left(\mu_n^{t, u e}\right)^2\left(1-\lambda_n^t / \mu_n^{t, u e}\right)},
\end{align}
where the first and the second term represent the processing delay and queuing delay, respectively. $\mu_n^{t, u e}=(f_n^{t, u e}/\rho \sum_{l=0}^{\ell_n^t}M_n(l))$ is the service rate of UE $n$, and $\sum_{l=0}^{\ell_n^t} M_n(l)$ is the computational complexity (MACs) to accomplish the local portion of the inference of task $\phi_n$. $\rho$ (cycle/MAC) represents the number of CPU cycles required to complete a MAC, which depends on the CPU model \cite{guangjin}.

2) Transmission delay: We consider a frequency division multiple access (FDMA) method for channel access, which means that all UEs share the uplink bandwidth $W$ with allocation rate $\alpha_n^t$, $\alpha_n^t \in[0,1]$ and $\sum_{n \in \mathcal{N}} \alpha_n^t \leq 1$. We use $p_n$ and $h_n^t$ to denote UE $n$'s transmission power and channel gain, respectively. The achievable data rate of UE n can be expressed as
$R_n^t=\alpha_n^t W \log _2\left(1+\frac{p_n h_n^t}{\alpha_n^t W N_0}\right)$, and the transmission delay of the task $\phi_n$ can be represented as
\begin{align}
T_n^{t, trans}=\frac{\psi_n\left(\ell_n^t\right)}{R_n^t}
\end{align}

3) Edge sojourn delay: According to the serial queueing model in \cite{jointdnn}, the average sojourn delay on subsequent ES’s VM of UE can be obtained using the G/D/1 queueing model. However, if taking into account the substantial computational capability of ES, we can disregard the queuing delay of ES and express the edge sojourn delay as
\begin{align}
T_n^{t, e s}=\frac{1}{\mu_n^{t, e s}}
\end{align}
where $\mu_n^{t, es}=(f_n^{t, es}/\rho \sum_{l=\ell_n^t+1}^{L_n}M_n(l))$ is the service rate of ES's VM for task $\phi_n$.

\subsection{Energy and Memory Model}
In time slot $t$, The energy consumption of a UE includes the local computation and the offloading transmission. The local computation energy consumption of UE $n$ is given by
$
E_n^{t, comp} =\kappa \rho\left(f_n^{t, u e}\right)^2 \sum_{l=0}^{\ell_n^t} M_n(l) \lambda_n^t
$
where $\kappa$ is a device-specific coefficient. The offloading transmission energy consumption for UE $n$ is denoted as
$
E_n^{t, trans}=p_n T_n^{t, trans}\lambda_n^t
$. Consequently, the total energy consumption of UE $n$ to execute task $\phi_n$ in slot $t$ can be expressed as
\begin{align}
E_n^{t, ue}=E_n^{t, comp}+E_n^{t, trans}
\end{align}

In order to constrain the memory resource used to deploy the pre-trained DNN model, we denote the memory cost factor on UE $n$ and ES with $\gamma_n$ and $\gamma_{es}$, respectively. According to \cite{Howfast}, the memory cost of $\phi_n$ can be written as
\begin{align}
C_n^{t, tot}=\gamma_n &\sum_{l=0}^{\ell_n^t} C_n(l)+\max \left(\psi_n(l), \cdots, \psi_n\left(\ell_n^t\right)\right)+\nonumber\\
\gamma_{es} \sum_{l=\ell_n^t+1}^{L_n} &C_n(l)+\max\left(\psi_n\left(\ell_n^t+1\right),\cdots,\psi_n\left(L_n\right)\right)
\end{align}
where the summation terms in the equation represent the memory footprint of DNN model parameters, while the max function terms represent the memory footprint of intermediate activation values during the inference process. 

\section{Problem Formulation}

In this section, we develop an optimization problem targeting the minimization of average E2E delay across all UEs while managing average energy and memory constraints. Our objective is to optimize DNN partitioning, computational resource, and communication resources allocation within each slot for collaborative edge inference. Notably, this is done without assuming knowledge of future random channel conditions and task arrivals. Thus, the optimization problem is formulated as a multi-slot stochastic mixed-integer nonlinear programming (MINLP) problem $\mathcal{P}$1:
\begin{align}
\min _{\left\{\bm{\mathrm{\ell^{\mathrm{t}}},\mathrm{\upalpha^{\mathrm{t}}},\mathrm{f}^{\mathrm{t,ue}},\mathrm{f}^{\mathrm{t,es}}}\right\}_{t=1}^K} & \lim _{K \rightarrow \infty} \frac{1}{K} \sum_{t\in \mathcal{T}} \sum_{n\in\mathcal{N}} T_n^{t, E2E} \nonumber\\
\text { s.t. C1: } & \lim _{K \rightarrow \infty} \frac{1}{K} \sum_{t\in \mathcal{T}} E_n^{t, {ue}} \leq e_n, \enspace\forall n \nonumber\\
\text { C2: } & \lim _{K \rightarrow \infty} \frac{1}{K} \sum_{t\in \mathcal{T}} C_n^{t, tot} \leq \varepsilon_n,\enspace \forall n \nonumber\\
\text { C3: } & \sum_{n\in \mathcal{N}} f_n^{t, e s} \leq f^{max, e s},\enspace \forall n, t \nonumber\\
\text { C4: } & \sum_{n\in \mathcal{N}} \alpha_n^t \leq1, \enspace \forall n, t \nonumber\\
\text { C5: } & \alpha_n^t, f_n^{t, e s}\geq 0, \enspace \forall n, t \nonumber\\
\text { C6: } & 0 \leq f_n^{t, ue} \leq f^{max, ue}, \enspace \forall n, t \nonumber\\
\text { C7: } & \mu_n^{t, u e}-\lambda_n^t>0, \enspace \forall n, t \nonumber\\
\text { C8: } & \ell_n^t \in\mathcal{L}, \enspace \forall n, t
\end{align}
where we denote $\bm{\mathrm{\ell^{\mathrm{t}}}}=\left\{\ell_1^t, \cdots, \ell_N^t\right\}$, $\bm{\mathrm{\upalpha}^{\mathrm{t}}}=\left\{\alpha_1^t, \cdots, \alpha_N^t\right\}$, $\bm{\mathrm{f^{\mathrm{t,ue}}}}=\left\{f_1^{t, ue}, \cdots, f_N^{t, ue}\right\}$, $\bm{\mathrm{f^{\mathrm{t,es}}}}=\left\{f_1^{t, es}, \cdots, f_N^{t, es}\right\}$. 
 C1 and C2 limit the long-term energy consumption and memory costs, respectively, ensuring they don't exceed thresholds $e_n$ and $\varepsilon_n$. Constraints C3 and C6 impose limitations on computational resource at the ES and UEs, respectively. C4 limits the uplink bandwidth. C7 ensures that each UE's task arrival rate doesn't exceed its service rate, guaranteeing the stability of the local computation queue.

The main challenge in directly solving $\mathcal{P}$1 resides in decoupling multiple optimization variables across multiple time slots, while simultaneously satisfying the long-term inference cost constraints associated with $\mathcal{P}$1. 

\section{Lyapunov-guided Multi-Dimensional Optimization (LyMDO) algorithm}

To address problem $\mathcal{P}$1, we introduce a novel algorithm LyMDO. The algorithm operates in three main stages. Initially, we apply Lyapunov optimization to convert the original multi-slot stochastic MINLP problem $\mathcal{P}$1 into a per-slot deterministic problem, thereby enabling resolution within individual time slots. Subsequently, we employ DRL to ascertain partitioning decisions for online resolution. Lastly, the challenge of multi-resource allocation is addressed using convex optimization.

\subsection{Lyapunov Optimization}
To manage to cope with the long-term constraint C1 and C2, we respectively construct a set of virtual energy queues $\mathbf{Q}(t)=\{Q_n(t)\}_{n=1}^N$ and memory cost queues $\mathbf{W}(t)=\{W_n(t)\}_{n=1}^N$ for UEs. In particular, we initialize both $Q_n(0)$ and $W_n(0)$ to 0, and update them dynamically as
\begin{align}
&Q_n(t+1)=\left[Q_n(t)+\nu_e\left(E_n^{t, ue}-e_n\right)\right]^{+} \\
&W_n(t+1)=\left[ W_n(t)+ \nu_c\left(C_n^{t, tot}-\varepsilon_n\right) \right]^{+}
\end{align}
where $\nu_e$ and $\nu_c$ are positive scaling factor. These virtual queues indicate the deviation between the accumulated resources consumption of task $\phi_n$ and the unit constraint in the time slot $t$. They guide UEs and ES with dynamically adjusting policies to meet the constraints (C1) and (C2), ensuring their stability. We define the quadratic Lyapunov function which expresses the scalar measure of queue congestion as 
$
L(\mathbf{\Theta}(t)) \triangleq \frac{1}{2} \sum_{n=1}^{N}\left(Q_{n}(t)^{2}+W_{n}(t)^{2}\right)
$
where $\mathbf{\Theta }(t)=\{\mathbf{Q}(t), \mathbf{W}(t)\}$ is the total queue backlog. The corresponding Lyapunov drift can be expressed as 
$
\Delta L(\mathbf{\Theta}(t))=\mathbb{E}[L(\mathbf{\Theta}(t+1))-L(\mathbf{\Theta}(t)) | \mathbf{\Theta}(t)]
$
and its drift-plus-penalty function can be expressed as 
\begin{align}
\Delta_VL(\mathbf{\Theta}(t)) = \Delta L(\mathbf{\Theta}(t))+V\cdot\sum_{n=1}^{N}\mathbb{E}\{T_n^{t, E2E}| \mathbf{\Theta}(t)]
\end{align}
where $V>0$ is the weight parameter, which is used to balance the performance between drift and penalty. 

According to the drift-plus-penalty minimization approach 
\cite{twcom2}, we can minimize the objective function in $\mathcal{P}$1 while stabilizing the queue backlog $\mathbf{\Theta }(t)$ by means of minimizing $\Delta_VL(\mathbf{\Theta}(t))$. Especially, it can be proved.\footnote{The proof follows similar arguments as presented in \cite{twcom2} and is omitted here due to space constraints.} that the drift-plus-penalty has an upper bound expression. On the basis of the principle of opportunistic expectation minimization \cite{stochastic} and the ratiocination above, we can transform the multi-slot stochastic MINLP original problem $\mathcal{P}$1 to the following per-slot deterministic problem $\mathcal{P}$2 in each time slot $t$:
\begin{align}
\min_{\bm{\mathrm{\ell^{\mathrm{t}}},\mathrm{\upalpha^{\mathrm{t}}},\mathrm{f}^{\mathrm{t,ue}},\mathrm{f}^{\mathrm{t,es}}}}  \lim _{K \rightarrow \infty}\sum_{n\in\mathcal{N}}&\left(Q_{n}(t) E_{n}^{t, ue}+W_{n}(t) C_{n}^{t, tot}\right) \nonumber\\ 
&+V \cdot \sum_{n\in\mathcal{N}}T_{n}^{t, E2E} \nonumber\\
\text { s.t. C3-C8.}&
\end{align}

Through the above transformation, the problem $\mathcal{P}$2 can be solved in the current time slot without requiring any prior information. However, it should be noted that problem $\mathcal{P}$2 remains a NP-hard MINLP problem in each time slot. 

\begin{figure}
\centering 
\includegraphics[height=3in, width=3.26in]{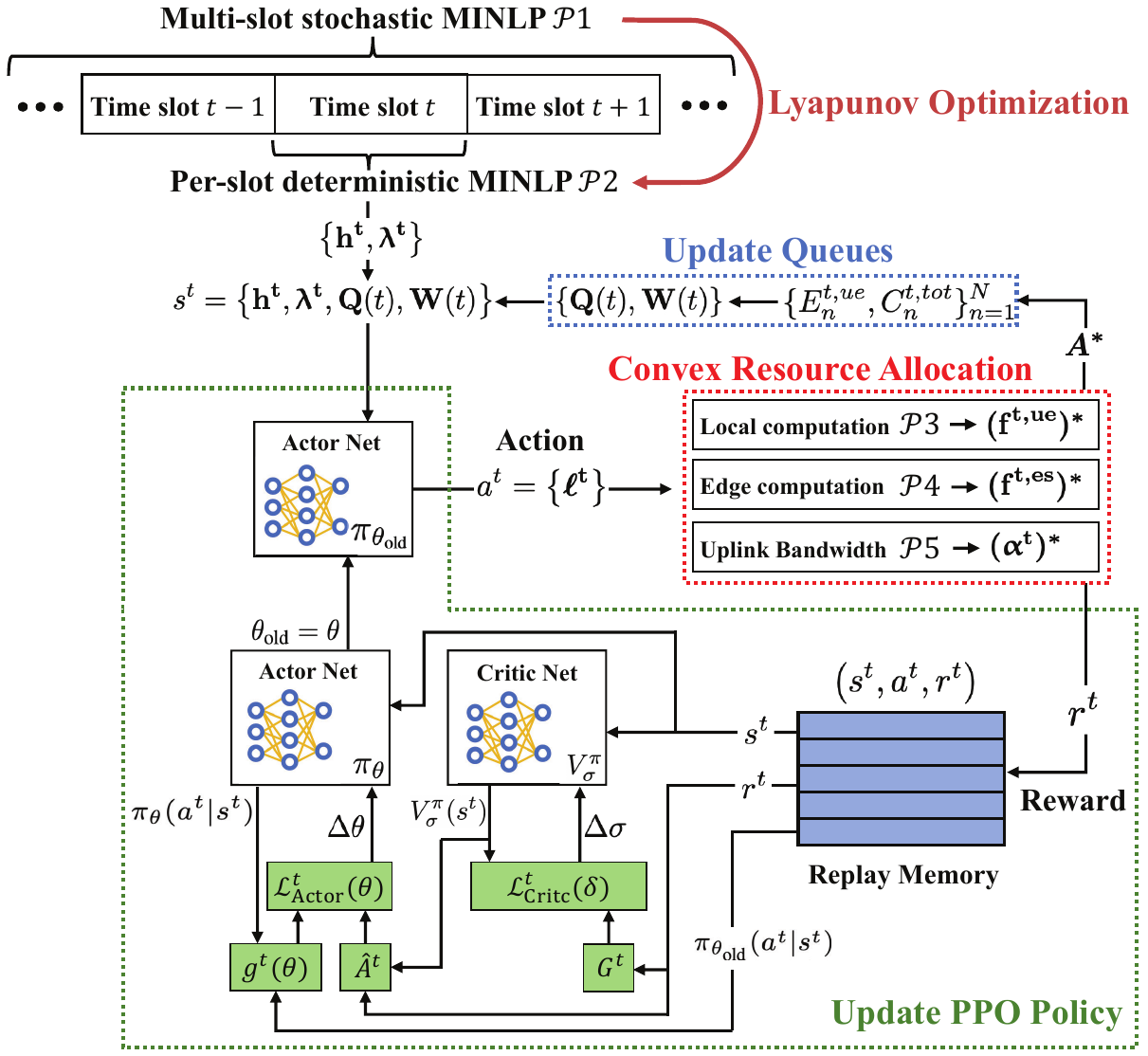} 
\caption{The schematics of LyMDO algorithm} 
\label{LyMDO} 
\end{figure}

\subsection{DRL Algorithm for Partitioning Decision}
To solve $\mathcal{P}$2, our objective is to design a online algorithm, which observes the system state information consisting of channel gains $\bm{\mathrm{h^{\mathrm{t}}}}=\left\{h_n^t\right\}_{n=1}^N$, task arrival rates $\bm{\mathrm{\uplambda^{\mathrm{t}}}}=\left\{\lambda_n^t\right\}_{n=1}^N$, and virtual queues $\left\{\mathbf{Q}(t), \mathbf{W}(t)\right\}$ at the beginning of each time slot to jointly generate the integer partitioning decision $\bm{\mathrm{\ell^{\mathrm{t}}}}$ and the continuous resources allocation decision $\left\{\bm{\mathrm{\upalpha^{\mathrm{t}}},\mathrm{f}^{\mathrm{t,ue}},\mathrm{f}^{\mathrm{t,es}}}\right\}$. This policy of LyMDO can be formulated as follows: 
\begin{align}
\bm{P}:\left\{\bm{\mathrm{h^{\mathrm{t}}}}, \bm{\mathrm{\uplambda^{\mathrm{t}}}}, \mathbf{Q}(t), \mathbf{W}(t) \right\} \rightarrow \bm{A^*}
\end{align}
where $\bm{A^*}=\left\{ \bm{\mathrm{\ell^{\mathrm{t}}}}, \left\{\bm{(\mathrm{\upalpha}^{\mathrm{t}})^*,(\mathrm{f}^{\mathrm{t,ue}})^*,(\mathrm{f}^{\mathrm{t,es}})^*}\right\} \right\}$. It involves mixed variables, and solving it directly using DRL would result in the issue of dimensionality explosion of action space. Upon reviewing $\mathcal{P}$2, we can observe that if we optimize
$\bm{\mathrm{\ell^{\mathrm{t}}}}$ and fix them first, the remaining resource allocation problems regarding $\left\{\bm{(\mathrm{\upalpha}^{\mathrm{t}})^*,(\mathrm{f}^{\mathrm{t,ue}})^*,(\mathrm{f}^{\mathrm{t,es}})^*}\right\}$ can be decoupled into three ``easy'' and separate convex optimization problems, which effectively reduces the difficulty of solving $\mathcal{P}$2.

Therefore, we decide to adopt proximal policy optimization (PPO), a policy-based DRL algorithm that exhibits notable advantages in terms of stability and reliability \cite{ppo}, to tackle DNN partitioning associated with variable $\bm{\mathrm{\ell^{\mathrm{t}}}}$. We commence by presenting our design of essential DRL components as follows:

1)\textbf{State}: The environment state of the DRL corresponds to the system state information of the aforementioned policy $\bm{P}$, so the state space can be represented as
$
s^t=\left\{\bm{\mathrm{h^{\mathrm{t}}}}, \bm{\mathrm{\uplambda^{\mathrm{t}}}}, \mathbf{Q}(t), \mathbf{W}(t) \right\}
$.

2)\textbf{Action}: To adapt the output of the actor network to the selection range $\mathcal{L}_n$ for different DNN networks, we employ the following mapping relationship between the output of the actor network $y_n^t$ and the partitioning decision $\ell_n^t$. Specifically, $tanh(\cdot)$ is employed as the activation function to transfer $y_n^t$ into $\left[-1,1 \right]$. The expression of mapping is written as
\begin{align}
\ell_n^t=\left\lfloor L_n \frac{\tanh (y_n^t)+1}{2}\right\rfloor
\end{align}
The action space in time slot $t$ can be expressed as
$
a^t=\left\{ \bm{\mathrm{\ell^{\mathrm{t}}}} \right\}
$

\begin{algorithm}
    \caption{DRL network training for LyMDO}
    \label{alg:LyMDO}
    \KwIn{Parameters $\left\{e_n, \varepsilon_n, \gamma_n \right\}_{=1}^K$, $\nu_e$, $\nu_c$, $V$, and other system parameters in $\mathcal{P}$1\;
          Maximum time slots $K$ and episodes $\delta$\;}
    \KwOut{Parameters $\theta$ of the convergent actor network\;}  
    Initialize the new actor network $\pi_{\theta}$ with $\theta$ and the critc network $V_\sigma^\pi$ with $\sigma$\;
    
    Initialize the old actor network $\pi_{\theta_{\text{old}}}$ with $\theta_{\text{old}}=\theta$\;
    
    Initialize an empty replay memory $\mathcal{M}$\;
    
    \For{$i=1,2,\cdots,\delta$}
    {
      Initialize energy queues $Q_n(1)=0$ and memory queues $W_n(1)=0$ for $\{\mathbf{Q}, \mathbf{W}\}$\;
      \For{$t=1,2,\cdots,K$}
      {
        Observe system state $\left\{\bm{\mathrm{h^{\mathrm{t}}}}, \bm{\mathrm{\uplambda^{\mathrm{t}}}}, \mathbf{Q}(t), \mathbf{W}(t) \right\}$\;
        
        Generate partitioning action $a^t=\left\{ \bm{\mathrm{\ell^{\mathrm{t}}}} \right\}$ from $\pi_{\theta_{\text{old}}}$ and then fix it\;

        Calculate probability $\pi_{\theta_{\text{old}}} (a^t \mid s^t)$\;
        
        Utilize the Fibonacci search to obtain local computational resource allocation $\bm{(\mathrm{f}^{\mathrm{t,ue}})^*}$\;
    
        Calculate edge computational resource allocation $\bm{(\mathrm{f}^{\mathrm{t,es}})^*}$ using (21)\;

        Utilize the CVX tool to obtain communication resource allocation $(\bm{\mathrm{\upalpha}^{\mathrm{t}})^*}$\;

        Calculate reward $r^t$ by equation (14)\;

        Update queues $\{\mathbf{Q}(t), \mathbf{W}(t)\}$ with $\bm{A^*}$\;

        Store $\left(s^t,a^t,r^t\right)$ in replay memory $\mathcal{M}$\;
        \If{$\mathcal{M}$ is filled}
        {
          \For{\rm{each sample in $\mathcal{M}$}}
          {
           Calculate $G(t)$ by equation (17)\;
           Generate partitioning action $a^t=\left\{ \bm{\mathrm{\ell^{\mathrm{t}}}} \right\}$ from $\pi_{\theta}$\;
           Calculate probability $\pi_\theta(a^t \mid s^t)$\;
           Calculate $V_\sigma^\pi(s^t)$ using critic network\;
           Calculate $\hat{A}^t$ by equation (17)\;
          }
          Replace $\theta_{\text{old}}$ of $\pi_{\theta_{\text{old}}}$ with $\theta$\;
          Update $\theta$ of $\pi_{\theta}$ and $\sigma$ of $V_\sigma^\pi$ according to (15) and (18)\;
          Clear replay memory $\mathcal{M}$\;
        }

      }
    }
\end{algorithm}

3)\textbf{Reward}: 
In our designed algorithm, Considering the minimization of the
objective function in $\mathcal{P}$2 while the PPO algorithm is designed to maximize the total discounted return, we set the negative objective function in $\mathcal{P}$2 as the reward in time slot $t$:
\begin{align}
r^t=-\sum_{n\in\mathcal{N}} \left(Q_{n}(t) E_{n}^{t, ue}+W_{n}(t) C_{n}^{t, tot}+ VT_{n}^{t, E2E} \right) 
\end{align}

As shown in Fig. 2, the core module of the multi-dimensional optimization algorithm is PPO. During its iterative process, the PPO policy is used to handle action $a^t$ of partitioning decisions. Then, the convex optimization module, which plays a crucial role in the PPO iteration process, is employed to solve the three subproblems related to resource allocation under the given $a^t$. Once the convex optimization problems are solved, $\bm{A^*}$ is applied to the environment, resulting in the update of energy and memory queues and the transition from state $s^t$ to $s^{t+1}$. Additionally, the reward $r^t$ is obtained and stored in the replay memory of PPO.

In the framework of PPO algorithm, the old actor network $\pi_{\theta_{\text{old}}}$ is responsible for generating the action $a^t$ based on the current state $s^t$. Additionally, a critic network $V_\sigma^\pi$ is utilized to evaluate the reward $r^t$ of $s^t$. ``state, action, reward'' in each time step are stored in the replay memory. During the training process, they are retrieved as samples to train the new actor network $\pi_{\theta}$ and $V_\sigma^\pi$. The widely used clipped surrogate loss \cite{ppo} to train $\pi_{\theta}$ is expressed as follows:
\begin{align}
\mathcal{L}_{\text{Actor}}^t(\theta)=\hat{\mathbb{E}}^t\left[\min \left[g^t(\theta) \hat{A}^t, \operatorname{clip}(g^t(\theta), 1-\epsilon, 1+\epsilon) \hat{A}^t\right]\right]
\end{align}
where $\mathbb{E}(\cdot)$ represents the expectation over a set of the training sample batch, and $g^t(\theta)=\frac{\pi_\theta(a^t \mid s^t)}{\pi_{\theta_{\text{old}}} (a^t \mid s^t)}$ is the ratio of the difference of policy between $\pi_{\theta}$ and $\pi_{\theta_{\text{old}}}$.
The truncation function (clip) is used to limit the magnitude of policy changes during each update. $\hat{A}^t$ is the estimated advantage value which can be expressed as
\begin{align}
\hat{A}^t=\sum_{i=0}^{K-t} \gamma^i \left(r^t+\gamma V_\sigma^\pi(s^t+1) -V_\sigma^\pi(s^t)   \right)
\end{align}
where $V_\sigma^\pi(s^t)$ is the output of critic network $V_\sigma^\pi$, and also the estimation of the discounted return which is represented as 
\begin{align}
G^t = \sum_{j=0}^{K-t} \gamma^jr^{t+j}
\end{align} 
Generally, the loss function used to train the critic network can be expressed as
\begin{align}
\mathcal{L}_{\text{Critic}}^t(\delta)=\left(V_\sigma^\pi(s^t)-G^t\right)^2
\end{align}

\subsection{Convex Optimization for Multi-Resource Allocation}
In this subsection, we introduce the utilization of three distinct convex optimization methods to solve three separate resource allocation problems and get $\left\{\bm{(\mathrm{\upalpha}^{\mathrm{t}})^*,(\mathrm{f}^{\mathrm{t,ue}})^*,(\mathrm{f}^{\mathrm{t,es}})^*}\right\}$, assuming that the partitioning decision $\bm{\mathrm{\ell^{\mathrm{t}}}}$ is given by $a^t$ of the PPO algorithm. 

1) Local computational resource:
For each UE in $\mathcal{N}$, there exists a local computation resource optimization subproblem $\mathcal{P}$3 as follows: 
\begin{align}
\min_{f_n^{t,ue}}  Q_n(t) \cdot \kappa_n&\left(f_n^{t, ue}\right)^2 d_n^{ue} \lambda_n^t \nonumber\\
+ V&(\frac{d_n^{ue}}{f_n^{t, ue}}+\frac{\left(d_n^{ue}\right)^2 \lambda_n^t}{2\left(\left(f_n^{t, ue}\right)^2-f_n^{t, ue} d_n^{ue} \lambda_n^t\right)})\nonumber\\
\text { s.t. C6, C7.}&
\end{align}
where $d_n^{u e}=\rho \sum_{l=0}^{\ell_n^t} M_n(l)$ due to the given $\ell_n^t$. The second derivative of the objective function in $\mathcal{P}$3 is consistently positive within its domain, thereby facilitating a straightforward proof of its convexity as an optimization problem. Considering that an analytical solution for $\bm{(\mathrm{f}^{\mathrm{t,ue}})^*}$ is difficult to obtain and $y_{ue}$ is a one-dimensional function, we employ the Fibonacci search, a line search method, to obtain the optimal value of it.

2) Edge computational resource:
The Edge computational  resource optimization subproblem $\mathcal{P}$4 can be expressed as
\begin{align}
\min_{\bm{\mathrm{f}^{\mathrm{t,es}}}}  V\sum_{n=1}^N \frac{ d_n^{e s}}{f_n^{t, e s}}\nonumber\\
\text { s.t. C3, C5.}
\end{align}
where $d_n^{es}=\rho \sum_{l=\ell_n^t+1}^{L_n} M_n(l)$ due to the given $\ell_n^t$. We can prove that $\mathcal{P}$4 is also a convex optimization problem due to the positive definite of its Hessian matrix. We can utilize the method of Lagrange multipliers to find its optimal solution, and the partial Lagrangian function is written as $\mathcal{L}_{\mathcal{P}4}$. By applying KKT conditions, we can derive the following necessary and sufficient conditions:
\begin{align}
&\frac{\partial \mathcal{L}_{\mathcal{P}4}}{\partial (f_n^{t, e s})^*}=-\frac{ Vd_n^{e s}}{f_n^{t, e s}}+u_{0}^{*}=0, (f_n^{t, e s})^*>0 \\
&u_{0}^{*}\left(\sum_{n \in \mathcal{N}^{*}} (f_n^{t, e s})^*-f^{max,es}\right)=0
\end{align}
Since both $\frac{Vd_n^{es}}{f_n^{t,es}}$ and $u_0^*$ are positive, we can combine (21) and (22) to obtain the analytical expression of $(f_n^{t,es})^*$ as
\begin{align}
\left(f_n^{t, e s}\right)^*=\frac{f^{max , e s} \sqrt{d_n^{e s}}}{\sum_{n\in\mathcal{N}} \sqrt{d_n^{e s}}}
\end{align}

3) Communication resource:
The communication resource optimization subproblem $\mathcal{P}$5 can be represented as
\begin{align}
\min_{\bm{\mathrm{\upalpha}^{\mathrm{t}}}}  \sum_{n\in\mathcal{N}}\left(Q_{n}(t) \cdot p_{n}\lambda_{n}^{t}+V\right)& \frac{\psi_n\left(\ell_n^t\right)}{\alpha_{n}^{t} W \log _{2}\left(1+\frac{p_{n} h_{n}^{t}}{\alpha_{n}^{t} W N_{0}}\right)}\nonumber\\
\text { s.t. C4, C5.}&
\end{align}
Numerous research papers on FDMA bandwidth allocation have demonstrated that the data rate, denoted as $\alpha_{n}^{t} W \log _{2}\left(1+\frac{p_{n} h_{n}^{t}}{\alpha_{n}^{t} W N_{0}}\right)$, is a convex function \cite{twcom2}. Furthermore, the fact that positive values over its entire domain serves as proof that its reciprocal, $1\bigg/ \left(\alpha_{n}^{t} W \log _{2}\left(1+\frac{p_{n} h_{n}^{t}}{\alpha_{n}^{t} W N_{0}}\right)\right)$, is also a convex function. Considering the constraints of $\mathcal{P}$5 are linear, we can utilize the CVX tool \cite{cvx} to solve it and obtain $\bm{(\mathrm{\upalpha}^{\mathrm{t}})^*}$.
The detailed pseudo-code for DRL training in LyMDO is presented in Algorithm 1.


\addtolength{\topmargin}{0.051in}

\section{Numberical Results}
\subsection{Simulation Setup}
In this section, we evaluate the performance of the proposed LyMDO algorithm through simulations. Unless otherwise stated, we set the system uplink bandwidth as $W=5$ MHz and the noise power spectral density as $N_0=174$ dBm/Hz \cite{twcom2}. We assume that the average channel gain $\bar{h}_n$ follows the free-space path loss model $\bar{h}_n=A_{d}\left(\frac{3 \times 10^{8}}{4 \pi f_{c} d_{n}}\right)^{d_{e}}$, $n\in\mathcal{N}$, where $A_d=3$ denotes the antenma gain, $f_c=915$ MHz denotes the carrier frequency, $d_e=3$ denotes the path loss exponent, and $d_n=150$ m denotes the distance between UE $n$ and AP. The uplink channel $h_n$ follows the Rayleigh fading channel model as $h_n=\beta\bar{h}_n$, where $\beta$ represents an independent exponential random variable with a unit mean. We consider UEs $N=5$ where two UEs execute type I tasks using  AlexNet while the remaining three UEs execute type II tasks using ResNet18. Other system parameters for simulation are listed in Table \ref{Table 1}.

\begin{table}[h]
\begin{center}
\caption{Simulation Parameters}
\label{Table 1}

    \begin{tabular}{|c|c|c|}
    \hline
        $V=10$ & $p_n=0.1$ W & $\lambda=\{0.5-2.5\}$ request/second  \\
    \hline
        $\rho=0.12$ & $\kappa=10^{-28}$ & $e_n$(Alexnet, Resnet) $=(40,60)$ mJ \\
    \hline
        $\gamma_n=0.2$ & $\gamma_{es}=0.8$ & $\varepsilon_n$ (Alexnet, Resnet) $=(100, 30)$ MB \\
    \hline
        $\nu_e=100$ & $\nu_c=10$ & $f^{max}$ (ue, es) $=(1.5, 15)$ GHz \\
    \hline    
    \end{tabular}
    
\end{center}
\end{table}

We implemented PPO using PyTorch 2.0. Empirically, we set both actor and critic networks to have 2 hidden layers with 128 and 64 neurons, respectively. The learning rate of them is set to $3\times10^{-4}$ with the training of Adam optimizer. We set the maximum episodes $\delta$ to 2000 and the maximum steps (maximum time slots) $K$ per episode to 200. The parameter $\epsilon$ used to limit policy updates in (16) is set to 0.2.

\begin{figure}
\centering 
\includegraphics[height=1.87in, width=3.3in]{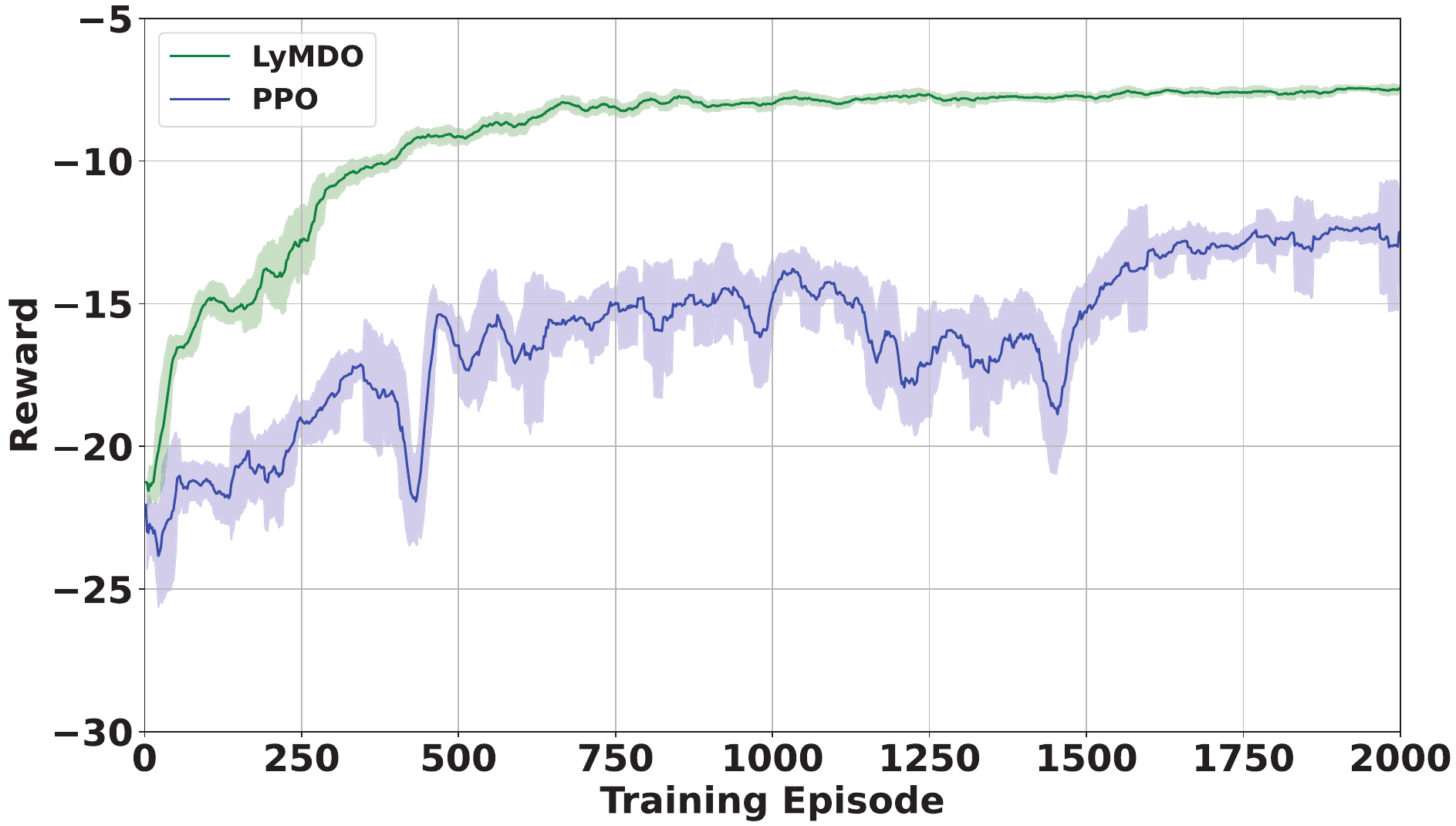} 
\caption{Comparison of convergence in the training process} 
\label{Convergence of Different Algorithms} 
\end{figure}

\subsection{Performance Evaluation}
In order to perform comparative experiments, we consider the following three baseline algorithms: 1)PPO: PPO to jointly optimize partitioning and resource allocation without the assistance of convex optimization \cite{Multi-exit}. 2)Local: Executing the entire task locally on UEs. 3)Edge: Offloading the entire task to ES for inference. 4)Random: The agent chooses the DNN partitioning decision randomly. It should be noted that the resource allocation for the last three baseline algorithms is optimized using our proposed convex optimization method.


In Fig. 3, we compare the convergence performance of the LyMDO and PPO algorithms. The reward curve of the PPO algorithm exhibits significant fluctuations and converges to a lower reward at 2000 episodes. This is because of the policy with a large action space ($4N$) the PPO algorithm needs to learn. Especially, the resource allocation strategy explored at the boundaries can result in long queuing delays or significant queue backlog to a sudden drop in reward. In contrast, the LyMDO algorithm only needs to learn the partitioning decision, while the remaining resource allocation is handled by convex optimization. This combined approach enables a stable improvement in reward during training, and it tends to converge around the 1000th episode.

\begin{figure}[htbp]
  \centering
  \begin{subfigure}[b]{0.45\linewidth}
    \centering
    \includegraphics[width=\linewidth]{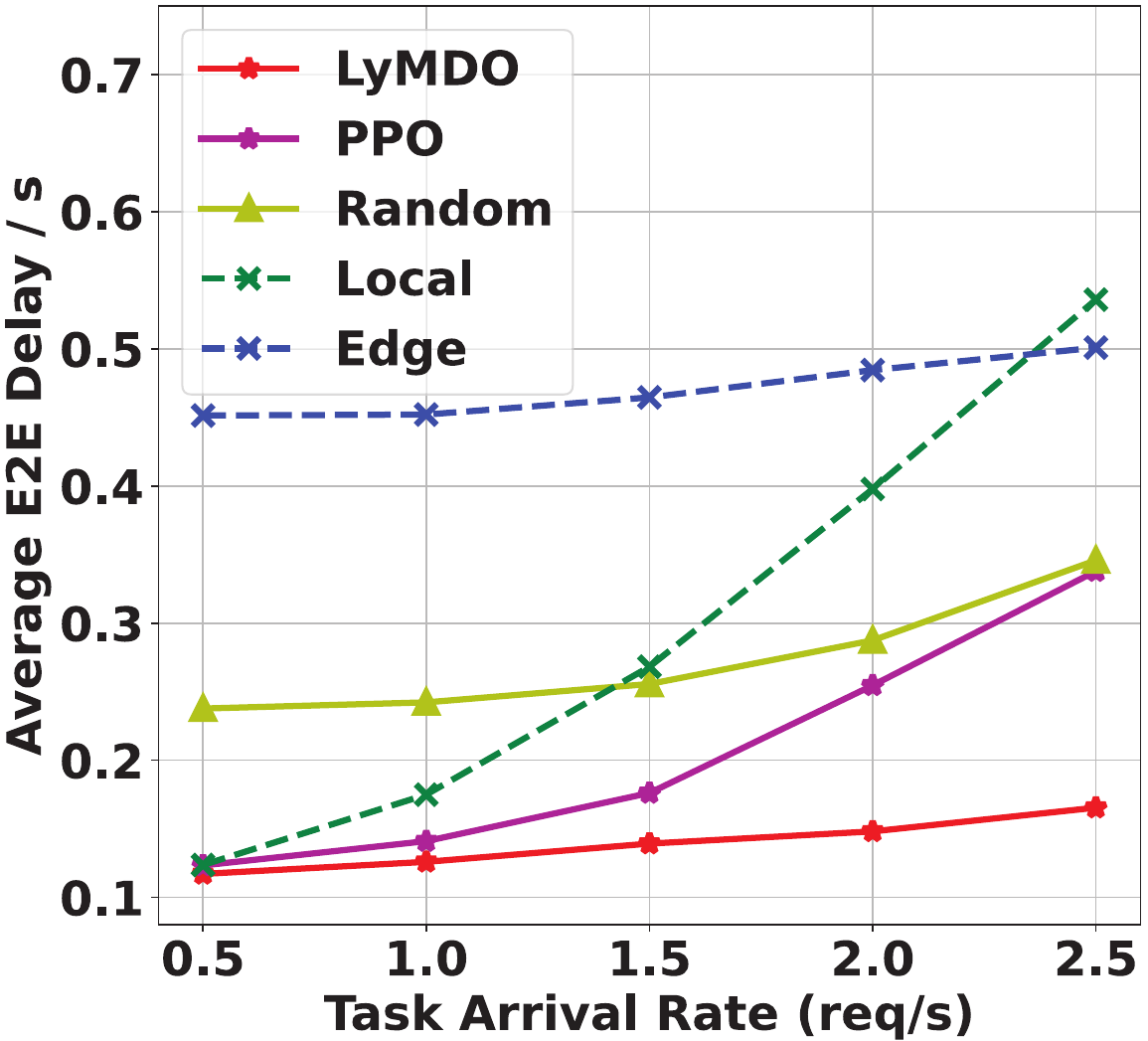}
    \caption{E2E delay}
    \label{fig:sub1}
  \end{subfigure}
  \hfill
  \begin{subfigure}[b]{0.45\linewidth}
    \centering
    \includegraphics[width=\linewidth]{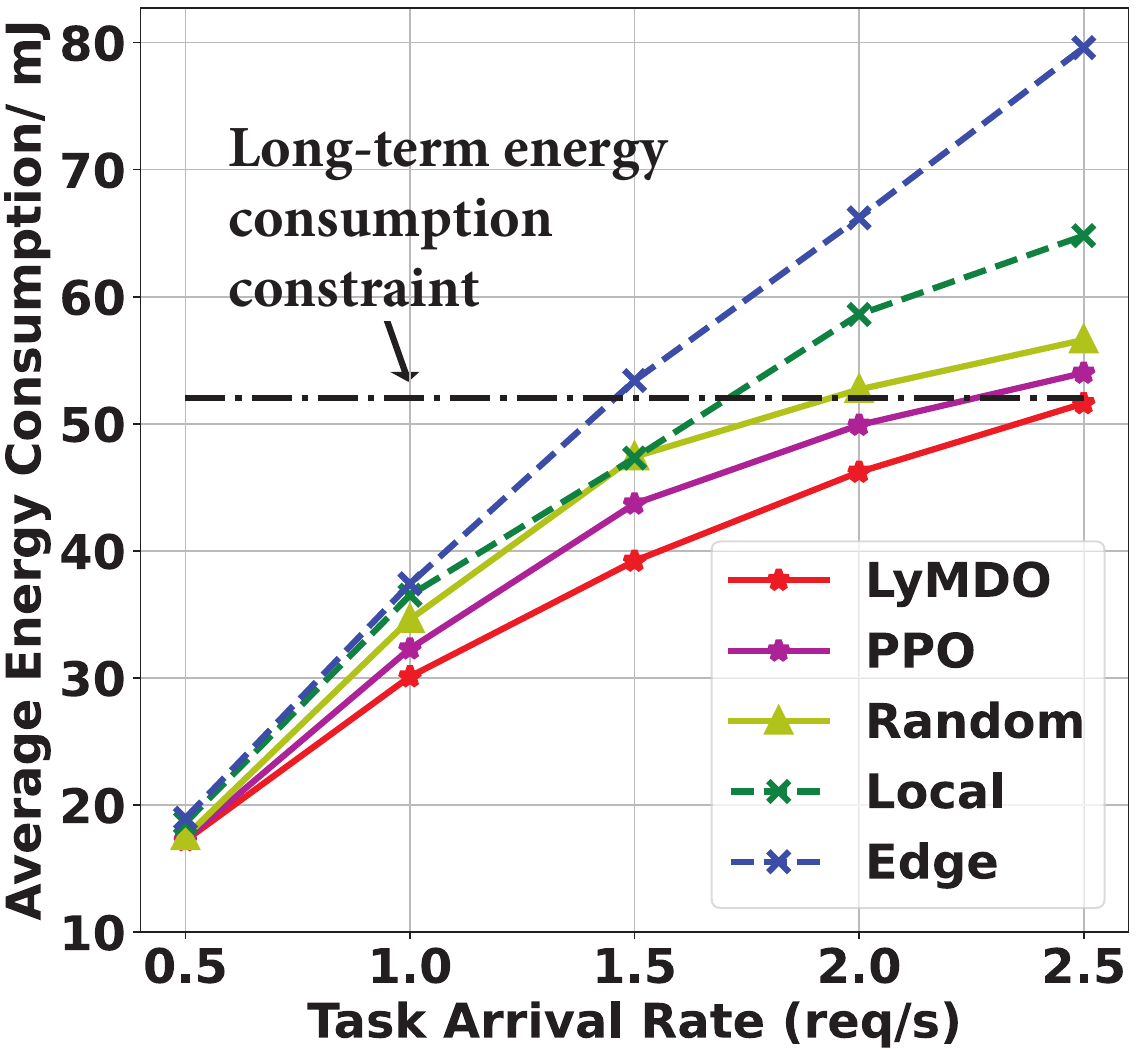}
    \caption{Energy consumption}
    \label{fig:sub2}
  \end{subfigure}

  \vspace{0.2cm} 

  \begin{subfigure}[b]{0.45\linewidth}
    \centering
    \includegraphics[width=\linewidth]{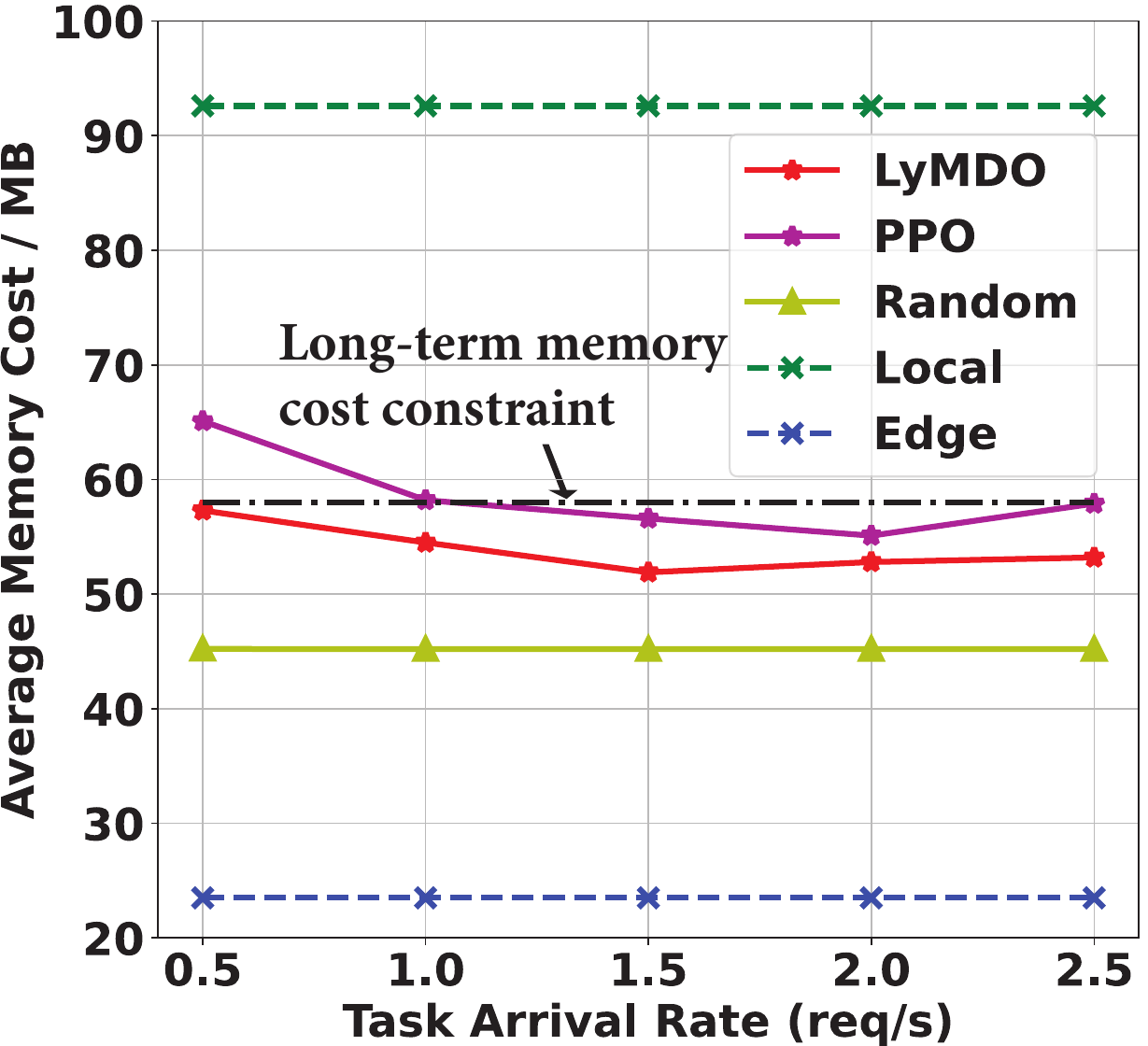}
    \caption{Memory cost}
    \label{fig:sub3}
  \end{subfigure}
  \hfill
  \begin{subfigure}[b]{0.45\linewidth}
    \centering
    \includegraphics[width=\linewidth]{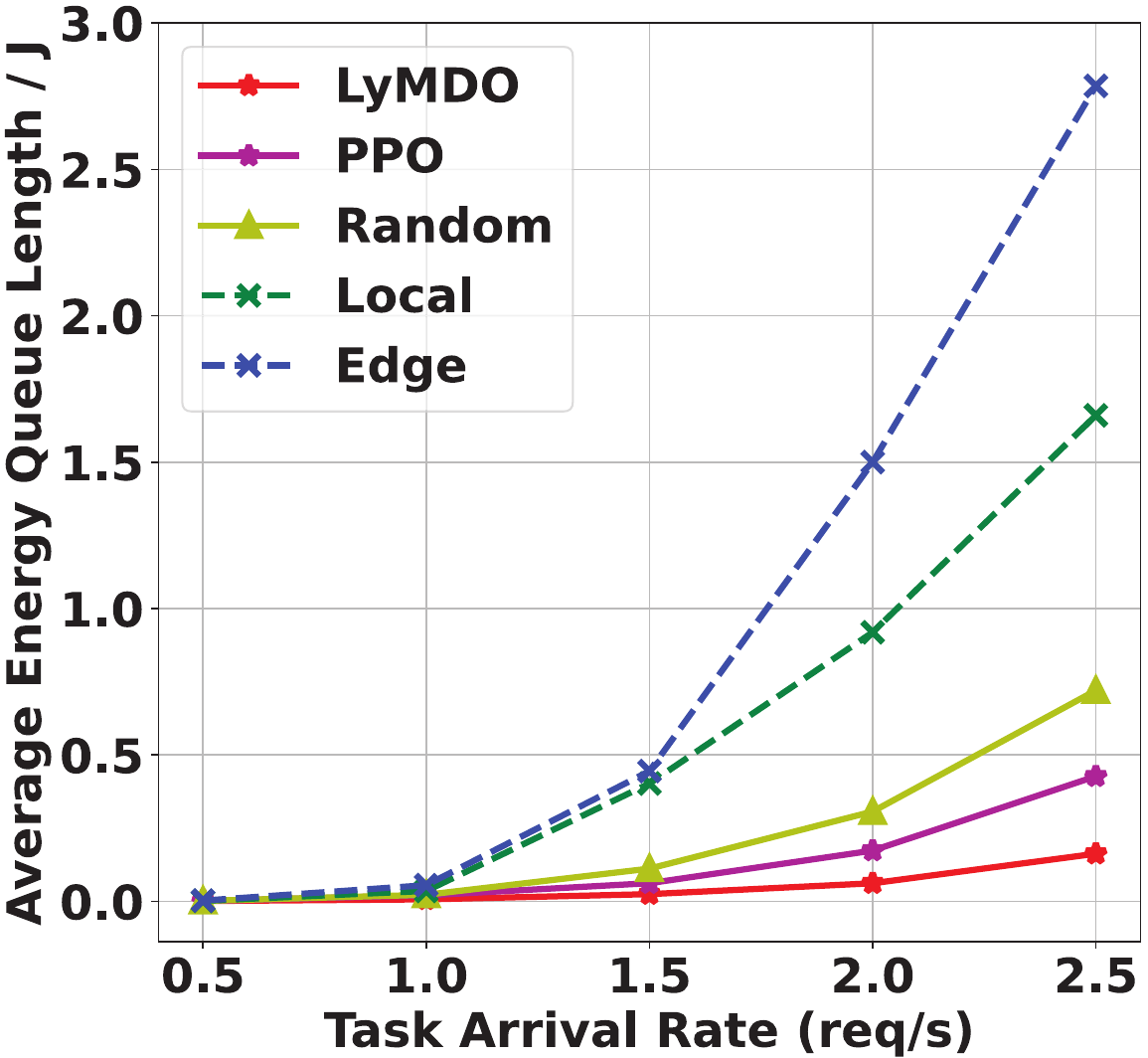}
    \caption{Energy queue length}
    \label{fig:sub4}
  \end{subfigure}
  \caption{Algorithms performance with respect to task arrival rates}
  \label{combined}
\end{figure}

\begin{figure}[htbp]
  \centering
  \begin{subfigure}[b]{0.49\linewidth}
    \centering
    \includegraphics[width=\linewidth]{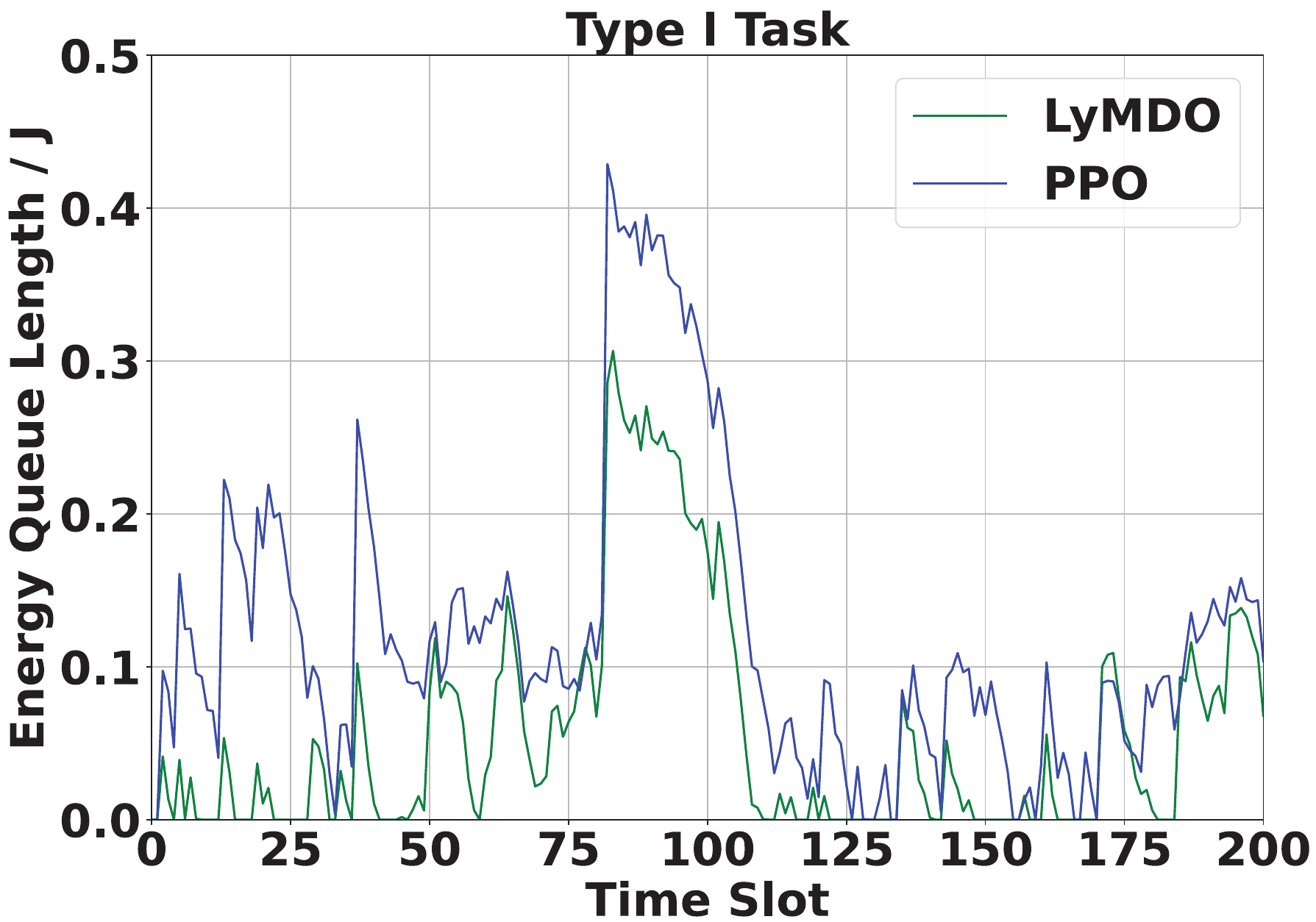}
    \caption{Task Type I with Alexnet}
    \label{sub1}
  \end{subfigure}
  \hfill
  \begin{subfigure}[b]{0.49\linewidth}
    \centering
    \includegraphics[width=\linewidth]{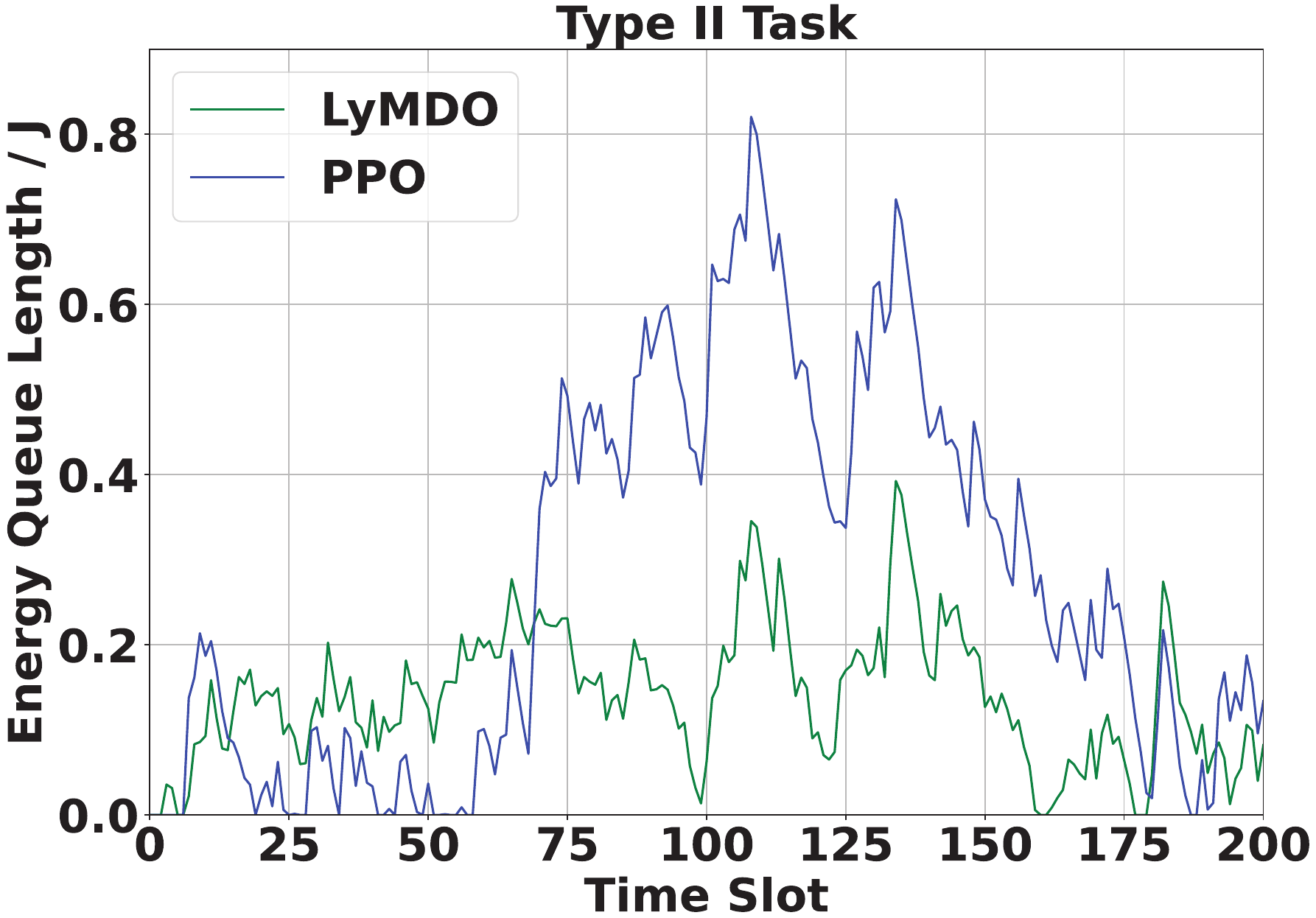}
    \caption{Task Type II with Resnet18}
    \label{sub2}
  \end{subfigure}
  \caption{Energy queue with respect to different task types}
  \label{fig:combined}
\end{figure}

In Fig. 4, After being well-trained offline, we compare the optimized performance of LyMDO with baseline algorithms under different arrival rates. Fig. 4(a) and (b) show that the average E2E delay and energy consumption increase with task arrival rates due to limited communication and computational resources in the MEC system. At low task arrival rates, the workload can be efficiently completed with local computation. As task arrival rates increase, LyMDO and PPO can adjust the partitioning decisions to utilize the multi-dimensional resources between ES and UEs for cooperative inference. The LyMDO achieves the lowest E2E delay and energy consumption, even under heavy workloads, while satisfying long-term energy consumption constraints. Fig. 4(c) demonstrates that LyMDO fully utilizes memory resource while satisfy the long-term memory cost constraints under proactive partitioning decision adjustments. The memory cost of the Random algorithm remains unchanged because a consistent random seed is applied across different arrival rates. Fig. 4(d) shows the robustness of LyMDO in maintaining a stable energy consumption queue under different workload conditions, due to the excellent environmental perception of LyMDO. Compared to PPO, LyMDO can reduce the average inference delay by approximately 30\%.

Fig. 5 presents a comparison of algorithms using the Lyapunov mechanism to stabilize the virtual queue for different types of DNN tasks. With the set to the task arrival rate of 2.5 (req/s), LyMDO demonstrates robust energy constraint capabilities for both ALexnet with low MACs and Resnet with high MACs. Particularly, during time slots 75 to 110 where task arrivals induce peak workload to result in energy queue backlog, LyMDO can promptly adjust the partitioning strategy, outperforming PPO. It effectively suppresses the continuous growth of queue length, reducing the energy queue length for both types of DNN tasks by 54$\%$ and 48$\%$, respectively.

\section{Conclusion}
This paper studies the long-term performance of a multi-user edge collaborative inference system. We employ a serial queue model to formulate the long-term E2E delay minimization problem under multi-resource constraints. To tackle it, we propose the LyMDO algorithm, which leverages a combination of PPO and convex optimization guided by Lyapunov optimization. This algorithm can individually perform online partitioning decisions and resource allocation in each time slot. The experimental results with different task arrival rates demonstrate that the proposed algorithm achieves lower E2E delay while satisfying all long-term constraints.

\section*{Acknowledgment}
This work was supported by the Innovation Program of Shanghai Municipal Science and Technology Commission under Grant 22511100604

\bibliography{reference.bib}
\end{document}